\documentstyle[12pt]{article}
\newcommand{\beq}{\begin{equation}}
\newcommand{\eeq}{\end{equation}}

\begin{document}
\begin{center}

{\bf Description of Dynamic Properties of Finite Electron Systems in
Density Functional Theory}\bigskip\bigskip

M. Ya. Amusia$^{a,b}$, V. R. Shaginyan$^{c,}$
\footnote{E--mail: vrshag@thd.pnpi.spb.ru}\bigskip

{\it $^{a}$The Racah Institute of Physics, Hebrew University,
Jerusalem 91904, Israel\\[0pt] $^{b}$A.F. Ioffe Physical-Technical
Institute, St.Petersburg 194021, Russia\\[0pt] $^{c}$Petersburg
Institute of Nuclear Physics, Gatchina 188350, Russia}\\
\end{center}\bigskip

\begin{abstract}
The self consistent version of the density
functional theory (DFT) is presented, which allows to calculate
the ground state and dynamic properties of finite multi-electron
systems such as atoms, molecules and clusters. The exact
functional equation for the effective interaction, using which one
can construct the action functional, density functional, the
response functions, and excitation spectra of the considered systems,
are discussed. We have also related the eigenvalues of the
single-particle Kohn-Sham equations to the real single-particle
spectra. We begin with the standard action functional and show that
it is useful in calculating the linear response functions $\chi$ via
which all physical characteristics of a many-body system can be
expressed.  It is shown, that function $\chi$ can be causal
(retarded), noncausal and advanced ones. This resolves the well known
paradox related to the causality and symmetry properties of the
response functions and the effective interaction.
\end{abstract} \bigskip\bigskip



\vspace{0.3cm}
\noindent {\bf INTRODUCTION}\bigskip

\noindent The density functional theory (DFT) originated from the
pioneer work of Hohenberg and Kohn \cite{wks} about thirty five
years ago. Since then an impressive progress has been achieved by
using DFT in describing microscopic many-particle systems and
especially in studying multi-electron objects, such as atoms,
molecules clusters and fullerenes. The recognition of DFT
successes culminated in awarding W. Kohn the Nobel Price in
Chemistry in 1998. At the beginning DFT was limited to
consideration of only the ground state properties of many-particle
systems, leaving aside their dynamic properties, which are closely
related to the system's behavior in the time-dependent external
fields. This limitation of DFT was overcome by Runge and Gross
\cite{rg} who thus have transformed DFT into the so-called
time-dependent density functional theory (TDDFT). Both, DFT and
TDDFT are based on the one-to-one correspondence between the
particle densities of the considered systems and external
potentials acting upon these particles. For the sake of simplicity
and definiteness, we assume that the time dependent part of the
considered electronic system's density $\rho({\bf r},t)$, which is
created under the action of an external time-dependent field
$\lambda v_{ext}({\bf r},t)$, is developing from the system's
ground state. As consequence of it one can conclude that each
observable, describing the system, can be written as a functional
of the density, among them are the functionals of the ground state
energy $E[\rho]$ and of the action $A[\rho]$ \cite{wks,rg}. Then
DFT and TDDFT establish an exact correspondence between an
interacting many-body system and a fictitious non-interacting
Kohn-Sham system \cite{wks,rg,gdp}. The main advantage of the
mapping is that Kohn-Sham system is described by a system of
Hartree-type  single particle equations, which is not too
difficult to solve. As a result, after solving these equations in
the case of an external time-independent potential, one can obtain
and predict, at least in principle, the atomic, molecular, cluster
and solid bodies binding energies, phonon spectra, activation
barriers, and etc (see e.g. \cite{wks}). The same is true in the
case of a time-dependent potential when the solution, generally
speaking, yields the single-particle and collective excitation
spectra, describes the behavior of a system in strong external
fields, including that of atoms in short laser pulses (see e.g.
\cite{gdp,ksk}). Unfortunately, the one-to-one correspondence
establishes only the existence of the functionals in principle,
leaving aside a very important question: how one can construct
them in reality. This is why the success of DFT and TDDFT strongly
depends upon the availability of good approximations for the
functionals. As such, an approach based on the local density
approximation and on numerical simulations of the ground state
energy of a homogeneous in density electron system was suggested
already in \cite{wks}. However the real molecules or atoms are not
homogeneous and numerical simulations are not universally good.
So, both justification and improvement of this approximation is
required. But attempts to do it run into hard problems. They had
to be overcome and it is highly desirable to find a systematic way
to construct the required functionals. On the other hand, an
important element of DFT, which is the action functional
$A[\rho]$, is suspected to be ill-defined because of a number of
contradictions in it. Perhaps the most important among them is the
contradiction, which came from the analysis of the causality in
TDDFT. It appeared, that this contradiction play a central role in
creating a number of other difficulties \cite{gdp,bg,l}.  In fact,
it signals the existence of a very serious paradox:  on one hand,
according to the Runge-Gross theorem \cite{rg} $A[\rho]$ is a well
defined functional, while on the other hand, the use of this
functional leads to fundamental difficulties.

Here we report some recent developments in eliminating this
paradox. As we shall see, the methods used in resolving it give a
good opportunity to construct equations defining the functionals
$A[\rho]$, $E[\rho]$ and related to them effective interaction and
the linear response function. Thus, this Comment will focus on
formulating such an approach to the self-consistent density
functional theory, that is based not only on the theorems of the
existence of the functionals but also on the exact equations for
them.

We will start with discussion of the paradox and show how it is
eliminated, then go to single particle potentials, present the
functional equations for the energy and action, derive the
expression for the effective inter- electron interaction and
illustrate the effectiveness of the suggested scheme.

It will be demonstrated that the presented approach is at the same
time rather simple, as compared to the usual, say, diagram
many-body technique, and offers a possibility to calculate
relatively easy all the required characteristics of any
multi-electron systems.\\

\noindent {\bf SYMMETRY AND CAUSALITY}\\

\noindent Let us start with the discussion of the action
functional $A[\rho]$, which is of key importance for the subject
of this paper, because it permits to express the reaction of the
many-body system under consideration to an external field. Due to
the existence of one-to-one correspondence between the external
potential $\lambda v_{ext}({\bf r},t)$ \cite{rg} and the
time-dependent density $\rho({\bf r},t)$, the external potential
determines also the time-dependent wave function $\phi[\rho](t)$
of the system in question. Therefore, the expectation value of any
quantum mechanical operator is a unique functional of the density
. This statement is also true for the action functional $A[\rho]$,
which is determined by the expression \cite{rg,gdp}, \beq
A[\rho]\; =\; \int_{t_i}^{t_f} dt\langle \phi(t)[\rho]\left|
i\frac{\partial}{\partial t}-\hat{H}(t)\right|
\phi(t)[\rho]\rangle. \eeq The values of $t_i$ and $t_f$ can be
chosen arbitrarily, we take $t_i=-\infty; t_f=+\infty$. The
time-dependent wave function $\phi(t)$, which corresponds to a
stationary point of the action functional (1), is a solution of
the Schr\"odinger equation, $$i\frac{\partial}{\partial
t}\phi(t)=\hat{H}(t)\phi(t),$$ with the hamiltonian, that we
present as a sum of two terms,
\beq\hat{H}(t)=\hat{H}+\hat{H_1}(t). \eeq Here
$\hat{H}=\hat{K}+\hat{V}$ acting alone describes the ground state
of the considered system, while $\hat{H_1}(t)$ represents the
external field and is given by the following expression: $$
\hat{H_1}(t)=\int\hat{\rho} ({\bf r})\lambda v_{ext}({\bf
r},t)d{\bf r}.$$ Here $\hat{\rho}({\bf r})$ is the electron
density operator. The operator $\hat{K}$ is the hamiltonian of
non-interacting particles and $\hat{V}$ represents the
two-particle Coulomb interaction. Upon using eqs. (1,2) the
functional $A[\rho]$ can be presented as, \beq
A[\rho]=A_1[\rho]-\int\lambda v_{ext}({\bf r},t) \rho({\bf r},t)d
{\bf r} dt,\eeq with $A_1=A|_{\lambda=0}$ being a universal
functional of the considered system, which is explicitly
independent upon the external field. Using the action functional
$A_1$, it is possible to construct the inverse of the linear
response function $\chi$, i.e. $\chi^{-1}$, \cite{ksk}, \beq
\chi^{-1}({\bf r}_1,t_1,{\bf r}_2,t_2) =\frac{\delta^2
A_1[\rho]}{\delta\rho({\bf r}_1,t_1) \delta\rho({\bf r}_2,t_2)}.
\eeq One needs to know the function $\chi$, because all
characteristics of a many-particle system are expressed via it.
From eq. (4) it might seem as self evident, that not only
$\chi^{-1}$ but also the direct linear response function $\chi$
defined by $\chi^{-1}({\bf r}_1,t_1,{\bf r}_2,t_2)$  must be a
symmetrical function of its space and time arguments
\cite{gdp,bg,l}. However, $\chi$, being symmetric under
interchange of $({\bf r}_1,t_1)$ and $({\bf r}_2,t_2)$, cannot be
a causal function, while  determined in the usual way
$\chi({\bf r}_1,t_1,{\bf r}_2,t_2)$ is known to be a retarded
(causal) function and, therefore, must vanish $t_2>t_1$
\cite{pp,ll}. Therefore it was concluded that eq. (4) is incorrect
\cite{gdp,bg,l}, because the symmetry and causality looked as
contradicting each other. To overcome this difficulty, it was
suggested that the problem is related to the boundary conditions
$\delta\phi(t_i)=\delta\phi(t_f)=0$ which one has to enforce on
the variation of the wave function while deriving the
time-dependent Schr\"odinger equation from eq. (1). It was argued,
that since the action functional is defined only on the
sufficiently narrow domain of $v$-representable (VR) wave
functions and densities, the condition $\delta\phi(t_f)=0$ cannot
be fulfilled. The $v$-representability means that the wave
functions and corresponding densities are solutions of the
Schr\"odinger equation with an external potential $v({\bf r},t)$
\cite{wks,mk}. Thus, a non-vanishing boundary term appears when
the variation $\delta A$ is made. Therefore, it was recommended to
search for another functional to avoid the mentioned above
causality-symmetry problems \cite{l}. But, as it was demonstrated
above, the existence of the action functional is guaranteed by the
Runge-Gross theorem \cite{rg}, which play the key role in
establishing TDDFT. As a result, the paradox has thrown doubt on
the reliability of TDDFT and it became crucially important to
resolve it.

In eliminating this paradox, we shall follow the papers
\cite{ksk,asl}. Consider a stable multi-electron system evolving
from its ground state $\phi_0$ with the density $\rho_0$ under the
influence of a weak external potential $\lambda v_{ext}$, that is
in the limit $\lambda\to0$.  This condition prevents the system
under consideration from a steady heating, ensuring the existence
of the linear response function $\chi$ as well as its inverse
$\chi^{-1}$ \cite{pp,ll}. For such a system all the densities and
wave functions in the vicinity of the ground state are VR
including those which satisfy the boundary conditions
\cite{mk,gd,l1}.  Therefore the correct time-dependent density can
be obtained from the Euler equation \cite{rg}, \beq \frac{\delta
A[\rho]}{\delta\rho({\bf r},t)}=0.\eeq It is worth to note, that
this equation is more general than the boundary condition. It
means that (5) can be correct even if $\delta\phi(t_f) \neq 0$.
Now we can use (5) to prove (4). According to (3), the Euler
equation (5) can be written in the following form, \beq
\frac{\delta A[\rho]}{\delta\rho({\bf r},t)} =\frac{\delta
A_1[\rho]}{\delta\rho({\bf r},t)} -\lambda v_{ext}({\bf r},t)=0.
\eeq To prove (4), we expand the solution of (6) in terms of the
density fluctuations $\delta\rho=\rho_1-\rho_0$, induced by the
external field: \beq\frac{\delta A_1}{\delta\rho}|_{\rho_0}
+\frac{\delta^2 A_1}{\delta\rho^2}|_{\rho_0}\delta\rho -\lambda
v_{ext}|_{\lambda\rightarrow 0}=0,\eeq where $\rho_1$ is the
density in the presence of $\lambda v_{ext}$. The first term in
(7) vanishes because of the stationary condition (5). Upon using
the definition of the linear response function
$\chi=\delta\rho/\delta\lambda v_{ext}$, the exact equation is
obtained \cite{ksk}, \beq \chi^{-1}({\bf r}_1,t_1,{\bf r}_2,t_2)=
\frac{\delta^2 A_1[\rho]}{\delta\rho({\bf r}_1,t_1)
\delta\rho({\bf r}_2,t_2)}|_{\lambda=0}. \eeq Thus, we come to the
conclusion that the second functional derivative of the action
integral $A_1$ appeared in eq. (8) defines and is defined by the
inverse of the linear response function, which, as it should be,
is a symmetrical function of its variables. To understand the
physical meaning of eq. (8), which defines a symmetrical function
$\chi^{-1}$, consider eq. (1) which determines the action $A$.  It
is known that by varying this equation with respect to $\phi^{*}$
one derives the Schr\"odinger equation for the retarded solution
$\phi$, $\phi_r({\bf r}, t)$, while the variation with respect to
$\phi$ leads to the advanced solution $\phi^{*}$,
$\phi_a^{*}({\bf r},-t)$. As a result,
the variation of the action functional with
respect to the density $\rho$ produces a superposition of the
advanced and retarded solutions. This is why eq. (8) defines a
symmetrical function. Now we are going to demonstrate that having
the noncausal function $\chi^{-1}$, it is possible to construct
any type of the linear response functions $\chi$: noncausal, and
causal (both retarded and advanced) linear response functions. The
response function $\chi$ is connected to its inverse by the
following relation: \beq \delta(t_1-t_3)\delta({\bf r}_1-{\bf
r}_3)= \int^{\tau_2}_{\tau_1}\chi^{-1}({\bf r}_1,{\bf
r}_2,t_1,t_2) \chi({\bf r}_2,{\bf r}_3,t_2,t_3)d{\bf r}_2dt_2,
\eeq We see that noncausal $\chi^{-1}$ is the kernel of the integral
equation (9). The character of solutions $\chi$ obtained from (9)
depends on the boundary conditions which are defined by numbers
$\tau_1$ and $\tau_2$. If the linear response function has to be
causal, that is to be nonzero only at $t_2\geq t_3$ then
$\tau_1=t_3$ and $\tau_2=\infty$. If noncausal $\chi$ is
required, then $\tau_1=-\infty$ and $\tau_2=\infty$, while in the
case of advanced $\chi$ it has to be $\tau_1=-\infty$, and
$\tau_2=t_3$. As we will see below, using a different way, the linear
response function may be chosen indeed causal, noncausal or advanced.
And vice versa, starting, for instance, with the causal linear
response function $\chi$, taken as the kernel of (9), one can
construct $\chi^{-1}$ of desirable kind using the same eq. (9) and
choosing the appropriate boundary conditions. Thus, we come to the
conclusion: it is the existence of the symmetrical (noncausal)
inverse $\chi^{-1}$ that leaves no room for the paradox
\cite{ksk,asl}. Note that the particular case when $\chi^{-1}$ is
noncausal while $\chi$ is causal has been studied \cite{hb} recently.

It is worth to consider in more details the case when both $\chi$ and
$\chi^{-1}$ are causal since one may suspect that there are no such
solutions of (9) \cite{hb}. To show that the causal function
$\chi^{-1}_r$ does exist we consider the well known definition of the
linear response function, \beq \delta\rho({\bf r}_1,t_1)=
\int_0^{t_1} \chi({\bf r}_1,{\bf r}_2,t_1,t_2)
\lambda \hat{q}({\bf r}_2)v(t_2)d{\bf r}_2dt_2.  \eeq
For simplicity we substitute the external field in
eq. (10) by $v_{ext}({\bf r},t)=\lambda\hat{q}({\bf r})v(t)$. Then,
eq. (10) transforms into Volterra's integral equation of the first kind,
which always has a solution, see e.g. \cite{vt}. Thus, we conclude
that the causal inverse $\chi^{-1}_r$ exists.

In order to demonstrate that one can use the causal (retarded),
advanced and noncausal linear response functions equally effective
let us consider the second functional derivative of the action
with respect to external field. Because of the one-to-one
correspondence between external potential
$\lambda v_{ext}({\bf r},t)$ and
time dependent density $\rho({\bf r},t)$ such a functional $A_1[v]$
exists. Then, taking into account that the inverse of the linear
response $\chi$ is given by $\delta\lambda
v_{ext}/\delta\rho=\chi^{-1}$ and using (8), one obtains,
\beq
\chi({\bf r}_1,t_1,{\bf r}_2,t_2)= \frac{\delta^2
A_1[v]}{\delta\lambda v_{ext}({\bf r}_1,t_1) \delta\lambda
v_{ext}({\bf r}_2, t_2)}|_{\lambda=0}. \eeq
It is useful to derive eq. (11) directly from the definition
(1). We restrict ourselves to an infinitesimally weak
time-dependent field putting $\lambda=0$ after completing the
calculations. Therefore the usual  many-body technique of diagrams
can be applied in calculating the variational derivatives with
respect to the external field \cite{agd,kb}.
The variation $\Delta A_1$ of the action $A_1[v]$ induced by the
time-dependent external field
is then defined as, \beq\Delta A_1= \int_{t_i}^{t_f}
dt\langle \phi(t)\left| i\frac{\partial}{\partial
t}-\hat{H}\right| \phi(t)\rangle =\int_{t_i}^{t_f}
dt\frac{<\phi_0|T(\hat{H}_{\tau 1}(t)S(\infty))|\phi_0>}
{<\phi_0|T(S(\infty))|\phi_0>},\eeq where $T$ is the usual
time-ordering operator, and $\hat{H}_{\tau 1}(t)$
is $\hat{H_1}(t)$ in
the interaction representation with respect to the hamiltonian
$\hat{H}$. The expression for
$S(\infty)$ is defined by the following way, \beq
S(\infty)=T\exp\left[-i\int_c\hat{H}_{\tau 1}(t)dt\right], \eeq where
$c$ is the contour of integration, which starts at $-\infty$ and ends
at $+\infty$. Note that the functional $A_1[v]$ formally defined
as, \beq A_1[v]=i\ln<\phi_0|T(S(\infty))|\phi_0>, \eeq will give
the variation $\Delta A_1$ with respect to the external field,
which coincides with the variation following from eq. (14).
Calculating the second variational derivative of $A_1$ given by
eq. (14) with respect to the external field $\lambda v_{ext}$ and
putting $\lambda=0$ in the final result, one gets the following
relation: \beq\frac{\delta^{2}A_1[v]}{\delta\lambda v_{ext}({\bf
r}_1,t_1)\delta \lambda v_{ext}({\bf r}_2,t_2)}\mid_{\lambda=0}
=\chi({\bf r}_1,t_1,{\bf r}_2,t_2) \eeq $$ =-i(<\phi_0\mid T
\rho({\bf r}_1,t_1)\rho({\bf r}_2,t_2)\mid\phi_0>-<\phi_0\mid
\rho({\bf r}_1,t_1)\mid\phi_0> <\phi_0\mid \rho({\bf
r}_2,t_2)\mid\phi_0>).$$ It is seen from eq. (15) that the second
functional derivative of the action integral $A_1$ with respect to
the external field is defined by the noncausal linear response
function, which, as it should be, is a symmetrical function of its
variables \cite{ksk}. After performing the Fourier transformation
with respect to time, $\chi({\bf r}_1,{\bf r}_2,\omega)$ takes the
form, \beq \chi=\sum_{k\neq 0}\left[\frac{<0\mid
\hat{\rho}^{\dagger}({\bf r}_1)\mid k><k\mid \hat{\rho}({\bf
r}_2)\mid 0>}{\omega-(E_k-E_0)+i\delta}\right.-\left. \frac{<0\mid
\hat{\rho}^{\dagger}({\bf r}_2)\mid k><k\mid \hat{\rho}({\bf
r}_1)\mid 0>}{\omega+(E_k-E_0)-i\delta}\right].\eeq Here $\mid
0>,\,E_0$ and $\mid n>,\,E_k$ are the many-particle system's exact
ground and excited state wave functions and energies,
respectively. It is evident from (16) that $\chi$ has poles in
$II$ and $IV$ quadrants of the complex $\omega$ plane. It contains
exactly the same information related to the properties of a system
as the causal linear response function, while both these functions
coincide on the imaginary axis of the plane, being related one to
another by analytical continuation \cite {ll,kb}. Just in the same
way the response functions of higher orders can be constructed.
So, having at hand the action $A_1[\rho]$, one can study the
ground and excited states of a system. Thus, eq. (8) is of key
importance in calculations of the excited states within the TDDFT
framework \cite{ksk}. Note, that eq. (8) was later verified in
\cite{pgg}. It is pertinent to point here that one can use the
non-equilibrium Green function theory \cite{k,rs} to construct
both the retarded (causal) linear response function $\chi_r$,
which has poles in the $III$ and $IV$ quadrants, and the advanced
linear response function $\chi_a$, with poles being in the $I$ and
$II$ quadrants. The only formal difference from the used above
technique is the appearance of the counter ordered evolution
operator $T_{cr}$ instead of the usual time-ordering operator $T$
\cite{rs}. Thus we can use eq. (14) to evaluate $A_1[v]$ replacing
$T$ by $T_{cr}$, in this case eq. (14) is modified as, \beq
A_1[v]=i\ln<\phi_0|T_{cr}(S(\infty))|\phi_0>,\eeq with $T_{cr}$
being the contour-ordering operator, which orders the operators
along the contour $cr$ according to the  position on the counter
of their time arguments. Then, upon using eq. (17), we have
instead of eq. (15), \beq \frac{\delta^{2}A_1[v]}{\delta\lambda
v_{ext}^2}\mid_{\lambda=0} =\chi_r({\bf r}_1,t_1,{\bf r}_2,t_2)
=-i\theta(t_1-t_2)<\phi_0\mid\left[\rho({\bf r}_1,t_1) \rho({\bf
r}_2,t_2)\right]\mid\phi_0>.\eeq Here the brackets $[...]$ denote
the commutator of the density operators. In the same way as
$\chi_r$ was constructed, one can get the advanced linear response
function $\chi_a$, choosing a special contour $ca$ along which the
contour-ordering operator $T_{ca}$ orders the operators. Thus,
different forms of linear response functions correspond simply to
different choices of the contours. This result is quite obvious:
if one marks the moment on the contour then it is possible to
construct the retarded (advanced) linear response function, using
eq. (17) for the action functional. If no moment of time is marked
along the contour, and we deal with the original action functional
given by eq. (14), we shall have the linear response function
$\chi$ defined by eq. (15) in accordance with the previous
discussion. Thus, we can conclude that functional $A$ is in fact
well defined, while eq. (4), which determines the inverse function
$\chi^{-1}$, opens the possibility to calculate the dynamic
properties of a system including its excitation spectra
\cite{ksk,as,pgg}.\\

\noindent {\bf SINGLE PARTICLE POTENTIAL}\\

\noindent Now let us construct the single particle potential
$v_s[\rho]({\bf r},t)$ of the time-dependent Kohn-Sham equations.
It is convenient to define the exchange-correlation functional
$A_{xc}$ \cite{rg,gdp},
\beq A_{xc}[\rho]\equiv A_0[\rho]-A_{1}[\rho]
-\frac{1}{2}\int \frac{\rho({\bf r}_1,t)\rho({\bf r}_2,t)}
{|{\bf r}_1-{\bf r}_2|}dt\,d{\bf r}_1d{\bf r}_2,\eeq
in atomic units, used throughout this paper. Here $A_0$ is the
functional of non-interacting Kohn-Sham particles,
$$ A_0[\rho]=\int_{t_i}^{t_f} dt\langle \Phi(t)[\rho]\left|
i\frac{\partial}{\partial t}-\hat{K}\right|
\Phi(t)[\rho]\rangle, $$
with $\Phi(t)[\rho]$ being the unique time-dependent Slater
determinant of the density $\rho$ \cite{rg,gdp}.
The effective interaction $R^{(0)}$ is defined as \cite{ksk,as,s1},
\beq R^{(0)}({\bf r}_1,t_1,{\bf r}_2,t_2)\equiv
\frac{\delta(t_1-t_2)}{|{\bf r}_1-{\bf r}_2|} +\frac{\delta^2
A_{xc}}{\delta\rho({\bf r}_1,t_1) \delta\rho({\bf r}_2,t_2)}.\eeq
Combining eqs. (8)
and (20), the effective interaction $R^{(0)}({\bf r}_1,t_1,{\bf
r}_2,t_2)$ is determined by the relation, \beq R^{(0)}({\bf
r}_1,t_1,{\bf r}_2,t_2) =\chi_0^{-1}({\bf r}_1,t_1,{\bf r}_2,t_2)
-\chi^{-1}({\bf r}_1,t_1,{\bf r}_2,t_2),\eeq where
$\chi_0^{-1}=\delta^2 A_0/\delta\rho^2$ is the inverse of the
linear response function of non-interacting Kohn-Sham particles.
In the same way, calculating the functional derivatives of higher
orders, one can obtain the effective interaction $R^{(l)},\,l\geq
2$, \beq R^{(l)}({\bf r}_1,t_1,{\bf r}_2,t_2,...{\bf
r}_{l+2},t_{l+2})= \frac{\delta^{l+2}A_{xc} [\rho]}
{\delta\rho({\bf r}_1,t_1)\delta \rho({\bf r}_2,t_2)...\delta
\rho({\bf r}_{l+2},t_{l+2})}\mid_{\lambda=0}.\eeq Then, taking the
Fourier transform of the effective interaction with respect to
time, making the analytical continuation to obtain the causality,
and upon carrying out the inverse Fourier transform, one can get
the causal effective interaction terms $\Re^{(l)}$. These terms
$\Re^{(l)}$ are members of a series, which define the causal
effective interaction $\Re[\rho]$, \beq \Re[\rho]({\bf
r}_1,t_1,{\bf r}_2,t_2) \eeq $$=\sum_{l\geq
0}\int\frac{1}{l!}\Re^{(l)}({\bf r}_1,t_1,{\bf r}_2,t_2,...{\bf
r}_{l+2},t_{l+2})\delta \rho({\bf r}_{3},t_3)...\delta \rho({\bf
r}_{l+2},t_{l+2}) d{\bf r}_{3}dt_{3}...d{\bf r}_{l+2}dt_{l+2}.$$
Having at hand $\Re$, one can construct the single-particle
potential $v_s[\rho]({\bf r},t)$ by integrating the equation,
which determines this potential \beq \frac{\delta v_s[\rho]({\bf
r}_1,t_1)}{\delta \rho({\bf r}_2,t_2)}= \frac{\delta
v_{xc}[\rho]({\bf r}_1,t_1)}{\delta \rho({\bf
r}_2,t_2)}+\frac{1}{|{\bf r}_1-{\bf r}_2|} =\Re[\rho]({\bf
r}_1,t_1,{\bf r}_2,t_2), \eeq with $v_{xc}$ being the
exchange-correlation potential. Eq. (24) can be integrated as
simple as one can integrate the Tailor expansion of a function,
\beq v_s[\rho]({\bf r}_1,t_1)=\int\Re^{(0)}({\bf r}_1,t_1,{\bf
r}_2,t_2)\delta \rho({\bf r}_2,t_2)d{\bf r}_2dt_2 \eeq $$+
\sum_{l\geq 1}\int\frac{1}{(l+1)!}\Re^{(l)}({\bf r}_1,t_1,{\bf
r}_2,t_2,...{\bf r}_{l+2},t_{l+2})\delta \rho({\bf r}_{2},t_2)...
\delta \rho({\bf r}_{l+2},t_{l+2}) d{\bf r}_2dt_2...d{\bf
r}_{l+2}dt_{l+2}. $$ Then, one can use the exact functional
equation for $R({\bf r}_1,{\bf r}_2,\omega,g)$ \cite{ksk,as} to
construct approximations to $v_s[\rho]$, \beq R({\bf r}_1,{\bf
r}_2,\omega,g) =\frac{g}{|{\bf r}_1-{\bf r}_2|}\eeq $$-\frac{1}{2}
\frac{\delta^2}{\delta\rho({\bf r}_1,\omega) \delta\rho({\bf
r}_2,-\omega)} \int\int_0^g\chi({\bf r}_1',{\bf r}_2',iw,g')
\frac{1}{|{\bf r}_1'-{\bf r}_2'|} d{\bf r}_1\,'d{\bf
r}_2\,'\frac{dw}{2\pi}\,dg'.$$ Here $R({\bf r}_1,{\bf
r}_2,\omega,g)$ is the effective interaction depending on the
coupling constant $g$ of the Coulomb interaction. The coupling
constant $g$ in eq. (26) is varied in the range $(0-1)$. When
$g=1$ one has $R^{(0)}=R(g=1)$. In eq. (26) the linear response
function $\chi({\bf r}_1,{\bf r}_2,\omega,g)$ is the
Fourier-image in $(t_1-t_2)$ of the linear response function. It
is obtained from eq. (21) through algebraic transformation
and by introducing the coupling constant $g$,
\beq \chi({\bf r}_1,{\bf r}_2,\omega,g)=
\chi_0 ({\bf r}_1,{\bf r}_2,\omega)+ \int
\chi_0 ({\bf r}_1,{\bf r}'_1,\omega) R({\bf r}\,'_1,{\bf
r}\,'_2,\omega,g) \chi({\bf r}'_2,{\bf r}_2,\omega,g) d{\bf
r}'_1 d{\bf r}'_2, \eeq with $\chi_0$ being the linear response
function of non-interacting Kohn-Sham particles, moving in the
single-particle time-independent field \cite{ksk,as}. It is
evident that the linear response function $\chi(g=1)$ goes to
$\chi$ as $g$ goes to 1.

The physical meaning of (25) is quite transparent: as it is seen
from (24,25), the functions $\Re^{(l)}$ are directly obtained by
calculating the corresponding functional derivatives of
$v_s[\rho]$ with respect to the density. This shows that the
single particle time-dependent potential contains a lot of
information about the system. And, vise versa, to construct
$v_s[\rho]$ one needs the same information. It is important to
note that $v_s[\rho]$, determined by eq. (25), is suitable to
describe the system in a strong external field due to the general
property, given by eq. (3), with the provision that the series in
eq. (25) is convergent.

It is instructive to consider eqs. (8,15) in the case when
external field $\lambda v_{ext}$ is not weak. Deriving (8,15), we
used the fact that the linear response function and its inverse
exist when $\lambda=0$. In the case when $\lambda=\lambda_0\neq0$,
the existence of both of these functions is not self-evident.
Moreover, it is quite probable that in the general case their
existence is impossible to prove. The behavior of a system in an
arbitrary (including the strength) time-dependent field may be
very complicated and unstable, for instance, due to a permanent
transfer of the energy from the external field to the system under
consideration. On the other hand, the non-equilibrium theory is
fitted to describe time-dependent steady states, it is those
states that are stable against small perturbations \cite{k}. But
stability against small perturbations means that there exist the
response function of the system under consideration and its
inverse function. So let us suppose the existence of both of these
functions. This can be correct provided the series in eq. (25) are
convergent. Under this condition we can apply the same
consideration that has been used when deriving eqs. (8,15): the
density of the system imbedded in the
external field is VR, then the densities
in the neighborhood are also VR \cite{gd}.
As the result, the action functional $A[\rho]$ is defined, and eq.
(5) is valid. One can also apply eq. (7) to prove (8), having in
mind that $\lambda\to\lambda_0$, while $\rho_0$ in the considered
case is the time-dependent density in the external field
$\lambda_0 v_{ext}$. On the other hand, we can use eq. (17)
evaluating $A_1(\lambda_0)$, which is correct in the presence of
an external field. The symbol $(\lambda_0)$ emphasizes that the
external field is not a weak one. As a consequence, we obtain the
retarded response function $\chi_r(\lambda_0)$, given by eq. (17).
The advanced response function $\chi_a(\lambda_0)$ is defined in
the same manner by choosing the contour $ca$. The contour, which
starts at $-\infty$ and ends at $+\infty$, will produce the
response function $\chi(\lambda_0)$. Thus, we are coming to the
conclusion that the standard definition of the action functional
(1), as well as eqs. (8,15,17,18) are also valid in the considered
case of a strong external field. Further investigation of the
behavior of a system in this case will be published elsewhere.\\

\noindent {\bf EQUATIONS}\\

\noindent In this section we briefly outline the derivation of the
functional equation for the exchange-correlation functionals $E_{xc}$
and $A_{xc}$ when system in question does not perturbed by an
external field. In that case one has \beq
E_{xc}[\rho]=A_{xc}[\rho]|_{\rho(r,\omega=0)},\eeq
since $A_{xc}$ is also defined in the static densities domain. The
exchange-correlation functional may be obtained from
\cite{ksk,as}: \beq A_{xc}= -\frac{1}{2}\int\left[ \chi({\bf
r}_1,{\bf r}_2,iw)+ 2\pi\rho({\bf r}_1)\delta(w)\delta({\bf
r}_1-{\bf r}_2)\right] \frac{g'}{|{\bf r}_1-{\bf r}_2|}
\frac{dg'}{g'}\;\frac{dw}{2\pi}d{\bf r}_1d{\bf r}_2. \eeq Eq. (29)
presents the well-known expression for the exchange-correlation
energy of a system, see e.g. \cite{ksk,pp,as}. The only thing we
need, in order to consider eq. (29) as describing $A_{xc}$, is an
ability to calculate the functional derivatives of $A_{xc}$ with
respect to the density. This, as it is seen from eq. (29), reduces to
the ability of calculating the functional derivatives of the linear
response function $\chi$ with respect to the density $\rho({\bf
r},\omega)$ which was developed in \cite{as,s2}. Then the linear
response function is given by (27), and the effective interaction
$R$ is defined by the exact functional (26). The single-particle
potential $v_{xc}$, being time-independent, is of the form
\cite{ksk,as}, \beq v_{xc}({\bf r})
=\frac{\delta}{\delta\rho({\bf r},t)}A_{xc}|_{\rho=\rho_0}. \eeq
Here the functional derivative is calculated at $\rho=\rho_0$ with
$\rho_0$ being the equilibrium density. By substituting (29) into
(30), it can be shown that the single particle potential $v_{xc}$
has the proper asymptotic behavior, $v_{xc}(r\to\infty)\to-1/r$,
\cite{as,ab}. The potential $v_{xc}$ determines the energies
$\varepsilon_i$ and wave functions $\phi_i$ of fictitious
Kohn-Sham particles, \beq \left(-\frac{\nabla^2}{2}+V_H({\bf
r})+V_{ext}({\bf r}) +v_{xc}({\bf r}) \right)\phi_i({\bf r})
=\varepsilon_i\phi_i({\bf r}), \eeq
that in turn form the linear response function $\chi_0$,
\beq
\chi_0({\bf r}_1,{\bf r}_2,\omega)=\sum_{i,k}
n_i(1-n_k)\phi^*_i({\bf r}_1)\phi_i({\bf r}_2)
\phi^*_k({\bf r}_2)\phi_k({\bf r}_1)
\left[\frac{1}{\omega-\omega_{ik}+i\eta}
-\frac{1}{\omega+\omega_{ik}-i\eta}\right], \eeq
and the real density of the system $\rho$, \beq
\rho({\bf r})=\sum_i n_i|\phi_i({\bf r})|^2. \eeq
Here $n_i$ are the occupation numbers, $V_{ext}$ contains all
external single-particle potentials of the system, say the Coulomb
potentials of the nuclei. Then, $V_H$ is the Hartree potential, $$
V_H({\bf r})=\int\frac{\rho({\bf r}_1)}{|{\bf r}-{\bf r}_1|} d{\bf
r}_1,$$ and $\omega_{ik}$ is the one-particle excitation energy
$\omega_{ik}=\varepsilon_i-\varepsilon_k$, while $\eta$ being an
infinitesimally small positive number. The described above
equations (29-31) and (26,27) solve the problem of constructing
the self-consistent DFT: one can calculate $A_{xc}$, the ground
state energy and excitation spectra of a system without resorting
to approximations for $A_{xc}$ based on additional and foreign to
the considered problem calculations such as Monte Carlo
simulations, or something of this kind. We note, that using these
approximations, one faces difficulties in constructing the
effective interaction of finite radius and the linear response
functions \cite{gdp,bg}. On the base on the suggested approach one
can solve these problems. For instance, in the case of a
homogeneous electron liquid it is possible to determined
analytically an efficient approximate expression $R_{RPAE}$ for the
effective interaction $R$, which essentially improves the
well-known Random Phase Approximation \cite{as,s1} by taking into
account the exchange of electrons properly, thus forming the Random
Phase Approximation with Exchange. The corresponding expression for
$R_{RPAE}$ is as follows
\beq R_{RPAE}(q,g,\rho)=\frac{4\pi
g}{q^2}+R_E(q,g,\rho),
\eeq
where \beq
R_E(q,g,\rho)=-\frac{g\pi}{p_F^2}\left[\frac{q^2}{12p_F^2}
\ln\left|1-\frac{4p_F^2}{q^2}\right|
-\frac{2p_F}{3q}\ln\left|\frac{2p_F-q}{2p_F
+q}\right|+\frac{1}{3}\right].
\eeq
Here the electron density $\rho$ is connected to the Fermi
momentum by the ordinary relation $\rho = p^3_F/3\pi^2$. Having at
hand the effective interaction $R_{RPAE}(q,g,\rho)$, one can
calculate the correlation energy $\varepsilon_c$ per electron of
the electron gas with the density $r_s$. The dimensionless
parameter $r_s=r_0/a_B$ is usually introduced to characterize the
density, with $r_0$ being the average distance between electrons,
and $a_B$ is the Bohr radius. The density is high, when
$r_{s}\ll1$.

\vspace*{0.5cm} Table 1.\hfill\parbox[t]{12cm}{Correlation energy
per electron in eV of an electron gas of density $r_s$. The Monte
Carlo results \cite{mc} $\varepsilon^M_c$ are compared with the
RPA calculations and with the results of \cite{s1}.
$\varepsilon_{RPA}$ denotes the results of RPA calculations, and
$\varepsilon_{RPAE}$ denotes the results of the calculations when
the effective interaction $R$ was approximated by $R_{RPAE}$
\cite{s1}.}

\begin{center}
\vspace{0.5cm}
\begin{tabular}{|r|l|l|l|}  \hline\hline
$r_s$ & $\varepsilon^M_c$ & $\varepsilon_{RPA}$ &
$\varepsilon_{RPAE}$
\\ \hline\hline
1 & -1.62 & -2.14 & -1.62 \\ \hline
3 & -1.01 & -1.44 & -1.02 \\ \hline
5 & -0.77 & -1.16 & -0.80 \\ \hline
10 & -0.51 & -0.84 & -0.56 \\ \hline
20 & -0.31 & -0.58 & -0.38 \\ \hline
50 & -0.16 & -0.35 & -0.22\\  \hline\hline
\end{tabular}
\end{center}
\vspace{0.5cm}

As can be seen from Table 1, the effective interaction
$R_{RPAE}(q,\rho)$ permits to describe the electron gas
correlation energy $\varepsilon_c$ in an extremely broad interval
of density variation. Note, that even at $r_s=10$ the mistake is
no more than 10\% as compared to Monte Carlo calculations, while
the result becomes almost exact at $r_s=1$ and is absolutely exact
when $r_s\to 0$ \cite{s1}.

Now let us calculate the single particle energies $\epsilon_i$,
that, generally speaking, do not coincide with the eigenvalues
$\varepsilon_i$ of eq. (31). We remark that the
eigenvalues do not make a physical sense, see e.g. \cite{wks}. To
calculate the single particle energies one can use the Landau
equation \cite{ll},
\beq \frac{\delta E}{\delta n_i}=\epsilon_i.\eeq
Eq. (36) can be used since, as it follows from eqs. (27,32,33), the
density and the linear response function depend upon the occupation
numbers. Thus, one can consider the ground state energy as a
functional of the density and the occupation numbers, \cite{s3}, \beq
E[\rho({\bf r}),n_i]= T_k[\rho({\bf r}),n_i]+ \frac{1}{2}\int
V_H({\bf r})\rho({\bf r})d({\bf r})+ \int V_{ext}({\bf r})\rho({\bf
r})d({\bf r})+ E_{xc}[\rho({\bf r}),n_i].\eeq Here $T_k$ is the
kinetic-energy functional of non-interacting Kohn-Sham particles. It
follows from eqs. (29,31,36,37) that the single-particle energies
$\epsilon_i$ can be presented by the following expression,
\beq \epsilon_i=\varepsilon_i-<\phi_i|v_{xc}|\phi_i>
-\frac{1}{2}\frac{\delta}{\delta n_i} \int \left[\frac{\chi({\bf
r}_1,{\bf r}_2,iw)+ 2\pi\rho({\bf r}_1)\delta(w)\delta({\bf
r}_1-{\bf r}_2)} {|{\bf r}_1-{\bf r}_2|}\right] \frac{dwdg'd{\bf
r}_1d{\bf r}_2}{2\pi}.
\eeq
Here, as it follows from (27), $\delta\chi/\delta n_i$ is given by
the equation, \beq \frac{\delta\chi}{\delta n_i}=
\frac{\delta\chi_0}{\delta n_i}+ \frac{\delta\chi_0}{\delta n_i}R
\chi+\chi_0\frac{\delta R}{\delta n_i} \chi+\chi_0
R\frac{\delta\chi}
{\delta n_i}.\eeq
In (39) for the sake of brevity we omit the spatial integrations.
The variational derivative $\delta\chi_0/\delta n_i$ has the
simple functional form,
\beq \frac{\delta\chi^0({\bf r}_1,{\bf r}_2,\omega)}
{\delta n_i} =\left[G_0({\bf r}_1,{\bf r}_2,\omega
+\varepsilon_i) +G_0({\bf r}_1,{\bf r}_2,-\omega
+\varepsilon_i)\right] \phi^{*}_i({\bf r}_1)
\phi_i({\bf r}_2),\eeq
with $G_0({\bf r}_1,{\bf r}_2,\omega)$ being the Green function of
$N$ non-interacting electrons moving in the single particle
potential $V_H+v_{xc}+V_{ext}$.
We believe and some evidences can be
found in \cite{ksk} that the contribution coming from $\delta
R/\delta n_i$ is small. Nonetheless even having omitted this term,
we are still to deal with rather involved eqs. (38,39).
Therefore, it is of
interest to illustrate the general consideration of the one-electron
energies with a simple and important example, when only the exchange
part $A_{x}$ of the total exchange-correlation functional is selected
to be treated rigorously, using an approximation for
$A_{c}=A_{xc}-A_{x}$.  Taking into account eq. (28) and the local
density approximation, one gets,
\beq E_c[\rho]=\int\rho({\bf r})\varepsilon_c(({\bf r}))d{\bf r}.\eeq
While for functional
$A_{x}$ we have an exact expression \cite{as,s2}, \beq
A_{x}[\rho]=E_{x}[\rho]= -\frac{1}{2}\int\left[ \chi_0({\bf r}_1,{\bf
r}_2,iw) +2\pi\rho({\bf r}_1)\delta(w) \delta({\bf r}_1-{\bf
r}_2)\right] \frac{1}{|{\bf r}_1-{\bf r}_2|}
\frac{dw}{2\pi}d{\bf r}_1d{\bf r}_2. \eeq
As it is seen from eq. (30), single-particle potential $v_{xc}$
takes from,
\beq v_{xc}({\bf r})=v_{x}({\bf r})+v_{c}({\bf r}), \eeq
with the potentials being given by,
\beq v_{x}({\bf r})=\frac {\delta E_x}{\delta\rho({\bf r})};\,\,
v_{c}({\bf r})=\frac {\delta E_c}{\delta\rho({\bf r})}.\eeq
We remark that $v_x({\bf r})$ can be calculated exactly \cite{s2,gl},
while there are quite suitable approximations to the potential, see
e.g. \cite{gk}. Presenting single-particle potential $v_{xc}({\bf
r})$ in such a way, we simplify the calculations a lot, preserving
the mentioned above condition $v_{xc}(r\to\infty)\to-1/r$, which of
crucial importance for calculations of $\varepsilon_i$ related to the
Kohn-Sham unoccupied states, see eq. (31). In the same way as eq.
(38) was derived, one gets,
\beq \epsilon_i=\varepsilon_i-<\phi_i|v_{x}|\phi_i>
-\frac{1}{2}\frac{\delta}{\delta n_i} \int
\left[\frac{\chi_0({\bf r}_1,{\bf r}_2,iw)+
2\pi\rho({\bf r}_1)\delta(w)
\delta({\bf r}_1-{\bf r}_2)} {|{\bf r}_1-{\bf r}_2|}\right]
\frac{dw\,d{\bf r}_1d{\bf r}_2}{2\pi}.
\eeq
After straightforward calculations one arrives at rather
simple result for the single-particle spectra, that are to be
compared with experimental results,
\beq
\epsilon_i=\varepsilon_i-<\phi_i|v_{x}|\phi_i> -\sum_in_k\int
\left[\frac {\phi^*_i({\bf r}_1)
\phi_i({\bf r}_2) \phi^*_k({\bf r}_2)\phi_k({\bf r}_1)}
{|{\bf r}_1-{\bf r}_2|}\right]
d{\bf r}_1d{\bf r}_2. \eeq
The single-particle levels $\epsilon_i$, given by eq. (46),
resemble the ones that are obtained within the Hartree-Fock (HF)
approximation. If the wave functions $\phi_i$ would be solutions
of the HF equations and correlation potential $v_{c}({\bf r})$ were
be omitted, then the energies $\epsilon_i$ would exactly coincide
with their eigenvalues. But this is not the case since $\phi_i$ are
solutions of eq. (31), and the energies $\epsilon_i$ do not coincide
with either HF eigenvalues or with the ones of eq. (31).
We also anticipate that eq. (46) applying to solids will produce a
quite reasonable results for the energy gap at various high-symmetry
points in the Brillouin zone. \bigskip

\noindent {\bf CONCLUSIONS}\\

\noindent We have shown that it is possible to convert DFT into a
self-consistent theory by constructing the exact equation for the
exchange-correlation functional $A_{xc}$. The proposed version of
DFT has another desirable feature: a systematic way  to construct
successive approximations to $A_{xc}$ is becoming available. We
have shown that the usual action functional $A[\rho]$ can be used
to calculate the response functions. It was also confirmed that
the second functional derivative of the action functional with
respect to the density defines the linear response function. The
higher order functional derivatives define the higher order
response functions. All they are not the causal ones, being
symmetric under exchange of their arguments. We have also shown
that on the same ground one can calculate the retarded (causal)
and advanced response functions. Thus, the paradox related to the
symmetry properties of the effective interaction and the causal
properties of the linear response function is resolved. The
developed DFT allows for calculations of excitation spectra of any
multielectron system. We have also related the eigenvalues of the
single-particle Kohn-Sham equations to the real single-particle
spectrum.
\section*{\bf Acknowledgments}
This research was funded in part by the Russian Foundation for
Basic Research under Grant No. 98-02-16170 and INTAS under Grant No.
INTAS-OPEN 97-603.\\

\end{document}